\documentclass[conference]{IEEEtran}
\IEEEoverridecommandlockouts
\usepackage{cite}
\usepackage{amsmath,amssymb,amsfonts}
\usepackage{algorithmic}
\usepackage{graphicx}
\usepackage{textcomp}
\usepackage{xcolor}
\usepackage{multirow}

\usepackage{url}
\usepackage{tabularx}  

\usepackage{color}
\usepackage{tikz}
\usetikzlibrary{patterns}
\usetikzlibrary{calc}

\newcommand{\convexpath}[2]{
  [   
  create hullcoords/.code={
    \global\edef\namelist{#1}
    \foreach [count=\counter] \nodename in \namelist {
      \global\edef\numberofnodes{\counter}
      \coordinate (hullcoord\counter) at (\nodename);
    }
    \coordinate (hullcoord0) at (hullcoord\numberofnodes);
    \pgfmathtruncatemacro\lastnumber{\numberofnodes+1}
    \coordinate (hullcoord\lastnumber) at (hullcoord1);
  },
  create hullcoords
  ]
  ($(hullcoord1)!#2!-90:(hullcoord0)$)
  \foreach [
  evaluate=\currentnode as \previousnode using \currentnode-1,
  evaluate=\currentnode as \nextnode using \currentnode+1
  ] \currentnode in {1,...,\numberofnodes} {
    let \p1 = ($(hullcoord\currentnode) - (hullcoord\previousnode)$),
    \n1 = {atan2(\y1,\x1) + 90},
    \p2 = ($(hullcoord\nextnode) - (hullcoord\currentnode)$),
    \n2 = {atan2(\y2,\x2) + 90},
    \n{delta} = {Mod(\n2-\n1,360) - 360}
    in 
    {arc [start angle=\n1, delta angle=\n{delta}, radius=#2]}
    -- ($(hullcoord\nextnode)!#2!-90:(hullcoord\currentnode)$) 
  }
}

\def\BibTeX{{\rm B\kern-.05em{\sc i\kern-.025em b}\kern-.08em
    T\kern-.1667em\lower.7ex\hbox{E}\kern-.125emX}}
\begin{document}

\title{Coherence-driven inference for cybersecurity \\
}

\author{\IEEEauthorblockN{Steve Huntsman}
\IEEEauthorblockA{steve.huntsman@cynnovative.com}}

\maketitle

\begin{abstract}
Large language models (LLMs) can compile weighted graphs on natural language data to enable automatic coherence-driven inference (CDI) relevant to red and blue team operations in cybersecurity. This represents an early application of automatic CDI that holds near- to medium-term promise for decision-making in cybersecurity and eventually also for autonomous blue team operations.
\end{abstract}


\section{\label{sec:introduction}Introduction}

Red, blue, and purple team operations \cite{bryant2024ptfm} are central cybersecurity practices that are being reshaped by large language models (LLMs) \cite{singer2025feasibility,zhuo2025cyber}. However, LLMs are incapable of reasoning \cite{shojaee2025illusion} and require careful human oversight \cite{abuadbba2025promise}. Here, we indicate how to mitigate both shortcomings using a neurosymbolic architecture \cite{sarker2022neuro,marra2024statistical} that combines LLMs and \emph{coherence-driven inference} (CDI) \cite{thagard1989explanatory,thagard1998coherence,thagard2002coherence,blokpoel2024theoretical}.

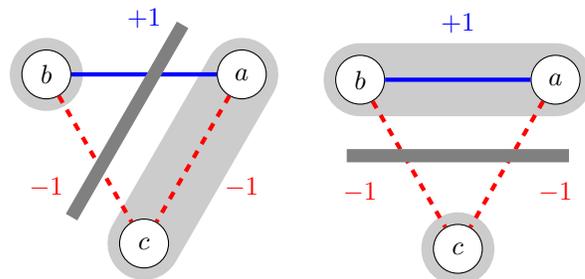
\begin{figure}[htbp]
    \centering   
        \begin{tikzpicture}[scale=1.0]
            \coordinate (A) at (30:1.5);
            \coordinate (B) at (150:1.5);
            \coordinate (C) at (270:1.5);
            \draw[thick,black!0,fill=black,opacity=0.2] \convexpath{A,C}{5mm};
            \draw[thick,black!0,fill=black,opacity=0.2](B) circle (5mm) node {};
            \node (q1) at (90:1.5) {{\color{blue}$+1$}};
            \node (q2) at (210:1.5) {{\color{red}$-1$}};
            \node (q3) at (330:1.5) {{\color{red}$-1$}};
            \node [draw,circle,fill=white,minimum size=6.5mm] (a) at (A) {$a$};
            \node [draw,circle,fill=white,minimum size=6.5mm] (b) at (B) {$b$};
            \node [draw,circle,fill=white,minimum size=6.5mm] (c) at (C) {$c$};
            \foreach \from/\to in {
                a/b}
                \draw[ultra thick, blue] (\from) to (\to);
            \foreach \from/\to in {
                a/c,b/c}
                \draw[ultra thick, red, dashed] (\from) to (\to);
            \coordinate (cut1) at (70:1.5);
            \coordinate (cut2) at (230:1.5);
            \draw[line width=5pt, color=gray] (cut1) to (cut2);
        \end{tikzpicture}
        \quad
        \begin{tikzpicture}[scale=1.0]
            \coordinate (A) at (30:1.5);
            \coordinate (B) at (150:1.5);
            \coordinate (C) at (270:1.5);
            \draw[thick,black!0,fill=black,opacity=0.2] \convexpath{A,B}{5mm};
            \draw[thick,black!0,fill=black,opacity=0.2](C) circle (5mm) node {};
            \node (q1) at (90:1.5) {{\color{blue}$+1$}};
            \node (q2) at (210:1.5) {{\color{red}$-1$}};
            \node (q3) at (330:1.5) {{\color{red}$-1$}};
            \node [draw,circle,fill=white,minimum size=6.5mm] (a) at (A) {$a$};
            \node [draw,circle,fill=white,minimum size=6.5mm] (b) at (B) {$b$};
            \node [draw,circle,fill=white,minimum size=6.5mm] (c) at (C) {$c$};
            \foreach \from/\to in {
                a/b}
                \draw[ultra thick, blue] (\from) to (\to);
            \foreach \from/\to in {
                a/c,b/c}
                \draw[ultra thick, red, dashed] (\from) to (\to);
            \coordinate (cut1) at (-10:1.5);
            \coordinate (cut2) at (190:1.5);
            \draw[line width=5pt, color=gray] (cut1) to (cut2);
        \end{tikzpicture}
    \caption{
    Left: the coherence graph $G$ associated with the propositions $a: X \text{ is hot}$, $b: X \text{ is bright}$, and $c: X \text{ is cold and dark}$. {\color{blue}Consistent (solid blue)} and {\color{red}inconsistent (red dashed)} pairs of vertices (= propositions) respectively get {\color{blue}positive} and {\color{red}negative} weights. Also shown is the cut $\{\{a, c\},\{b\}\}$ whose coherence is $-(A_{ab} + A_{bc}) = -(({\color{blue}+1}) + ({\color{red}-1})) = 0$, where $A$ is the weighted adjacency matrix of $G$. Right: the associated coherence graph $G$ with the cut $\{\{a, b\},\{c\}\}$ whose coherence is $-(A_{ac} + A_{bc}) = -(({\color{red}-1}) + ({\color{red}-1})) = 2$. CDI amounts to accepting and rejecting data according to a cut $\{U, V(G) - U\}$ that maximizes the coherence objective $-\sum_{u \in U, v \not \in U} A_{uv}$. The cut on the right is optimal, separating propositions according to their inconsistencies.
    }
    \label{fig:cdi}
\end{figure}

In CDI an individual datum such as a claim or proposition is represented by a vertex in a graph, as illustrated in Figure \ref{fig:cdi}.
Relevance and consistency relations between data are respectively represented by edges and weights in the interval $[-1,1]$. That is, data are compiled into a weighted \emph{coherence graph} that models which data relate to each other and how consistent they are with each other. Given a coherence graph $G$, CDI evaluates vertex bipartitions or \emph{cuts} that separate data into accepted and rejected sets. The objective is to find the cut that maximizes \emph{coherence}, defined as the negative sum of weights for edges that have one vertex in each part of the bipartition. Thus, maximizing coherence amounts to computing a maximum cut in a weighted graph: cuts with higher coherence tend to have more inconsistent edges and fewer consistent edges that cross from one part to the other. As an instance for the matrix $-A$ of the $\mathbf{APX}$-complete MAX-CUT problem, this is hard, but it can be solved exactly at useful scales and approximately at large scales \cite{khot2007optimal,moore2011nature,gartner2012approximation,lee2021classifying}. \footnote{In general, context matters for assigning acceptance and rejection to the results of CDI. Prioritizing direct observations and firm prior conclusions over secondhand reports and tentative prior conclusions is especially important.}

\begin{figure}[htbp]
    \centering
    \includegraphics[width=1\linewidth, trim={75 25 1075 25mm}, clip]{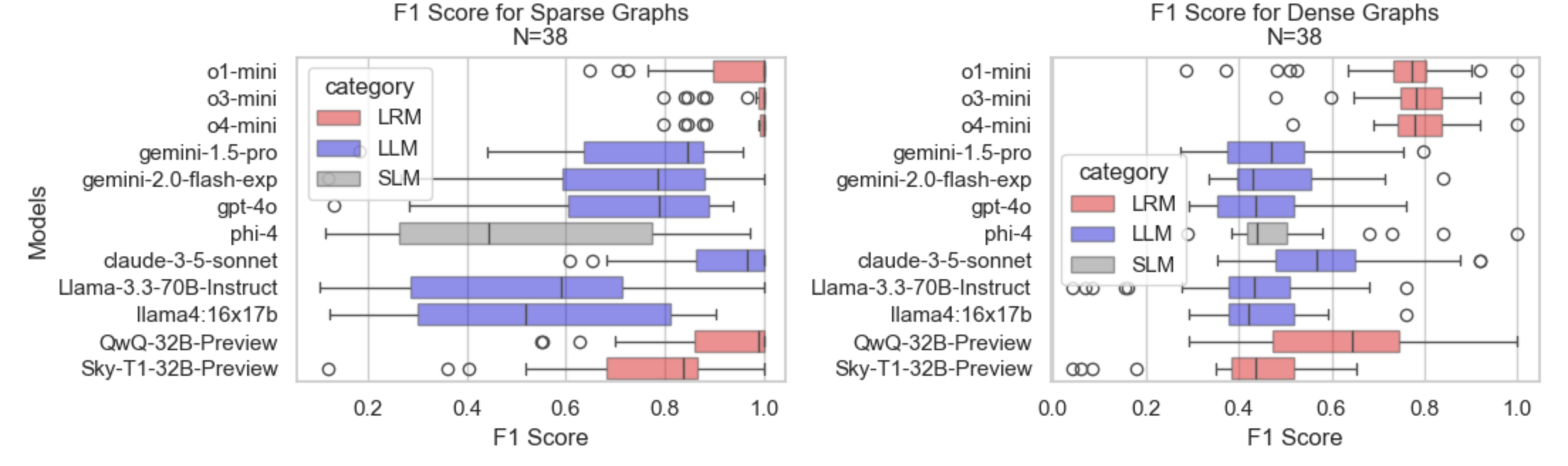}
        \caption{(Taken from \cite{huntsman2025neurosymbolic}, for which see details: LRM and SLM respectively indicate large reasoning and small language models.) Certain models have very high micro $F_1$ scores on sparse coherence graph reconstruction tasks.}
    \label{fig:cohere_graphs}
\end{figure}

An algorithmic benchmark demonstrates that powerful LLMs---particularly the o1/3/4 family of reasoning models---can accurately and reproducibly (re)construct coherence graphs from propositions using a single prompt \cite{huntsman2025neurosymbolic}. This suggests that CDI can be performed completely automatically. Figure \ref{fig:cohere_graphs} shows model performance on this benchmark. 

Accordingly, we use LLMs for ``fast'' or ``system 1'' reasoning with easy computations on rich representations \cite{lenat2023getting,bengio2024machine} and CDI for ``slow'' or ``system 2'' reasoning with hard computations performed on impoverished representations \cite{kahneman2011thinking}. Using an LLM to compile a coherence graph and symbolic reasoning via CDI to reason over the resulting graph naturally separates concerns and has many straightforward applications. It operates at a high level of abstraction, resolves ambiguity after forming representation, and is robust to parsing errors; it also uses the powerful representational ability of multi-head transformer architectures \cite{rajaraman2024transformers,liu2024infini} without relying on their weak computational abilities \cite{peng2024limitations}, though it can still benefit from the strengths of program synthesis and/or function calling if/as needed.

In this paper, we show examples of how automatically constructed median coherence graphs can inform red and blue team operations in cybersecurity. The idea is to extract (with a LLM model and/or manually) the most important propositions relating to a situation, then to compile a coherence graph over these claims using a LLM (in fact, doing this several times and taking a median of the vectorized adjacency matrix for robustness and stability\footnote{Throughout this paper, we take medians of $N = 15$ realizations of coherence graphs for robustness. Median edge weights are optimal in the sense of graph edit distance and also as an optimal rank one approximation with respect to the $L^1$ norm of a matrix whose columns are vectorized weighted adjacency matrices. As Figure \ref{fig:cdi1convergence} and similar subsequent figures show, considering medians of subsamples of increasing size also gives a way to gauge robustness.}), and finally to perform CDI on the result. 

Importantly, this use of LLMs appears to be performant and generally avoids hallucinations (see \S 2 of \cite{huntsman2024prospects} and \S B of \cite{huntsman2025neurosymbolic}). There is a simple explanation for this: the models essentially provide numerical scores summarizing how effectively they can \emph{interpolate} between pairs of propositions, instead of \emph{extrapolating} responses to more open-ended prompts. Moreover, our approach amounts to sampling independent identically distributed variables. Viewed in this light, a central limit theorem applies, and the only statistical concerns involve variance and convergence rate. We show these quantities throughout our discussion below, demonstrating that they behave well. In the event that this behavior is inadequate, sampling the relevant LLM chain of thought for relevant graph edges provides a transparent audit mechanism.

Efficiently, reliably, and reproducibly identifying incoherence in data and hypotheses could improve performance in both red and blue team operations. CDI has obvious applications to detecting, understanding, and combating cybersecurity threats. It may also lead to more capable autonomous blue team agents, but it is relatively unlikely to have the same impact on adversarial agent capabilities since extensive access to a powerful LLM is required.\footnote{
Note, however, that using a LLM as infrastructure for defensive cybersecurity agents presents its own class of vulnerabilities: we do not address these here.
}

CDI has many other applications to machine cognition (see, e.g., \cite{huntsman2025automatic} for another application that uses LLMs), but we focus here on a small set of illustrative examples. More practical applications can follow by combining suitable patterns and templates for sets of propositions relevant for red and especially blue team operations; modular graph construction and approximate solvers for improved scalability;
\footnote{The examples we provide throughout are actually already at scales similar to those in the literature on CDI. It is not clear that directly scaling to larger inferences is particularly useful, especially compared to structuring inference hierarchically and/or sequentially in closer correspondence to human cognition. Indeed, naive scaling to larger coherence graphs to achieve more stronger cognitive abilities is of questionable utility in light of a fixed-parameter tractable cognition hypothesis \cite{van2008tractable,van2019cognition}.}
and most importantly, experimentation by users in practical settings. In particular, CDI can and should be compared to and contrasted against more straightforward applications of LLMs to color team operations such as \cite{singer2025feasibility,zhuo2025cyber}.

The paper is organized as follows. Section \S \ref{sec:decisionMaking} gives a toy example and relates it to cybersecurity. Sections \S \ref{sec:blue} and \ref{sec:red} respectively address examples relevant to blue and red team operations before the conclusion in \S \ref{sec:conclusion} and an appendix detailing the basic prompt structure we use.

\section{\label{sec:wifi}A toy problem}

\begin{figure}[htbp]
\centerline{\includegraphics[width=.55\columnwidth, trim={0 0 0 0mm}, clip]{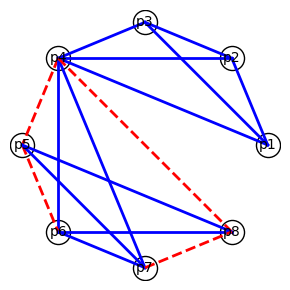}}
\caption{The median of 15 individual coherence graphs obtained using a single prompt to o1-mini to evaluate the propositions in Table \ref{tab:wifiPropositions}. The optimal cut here has smaller part $\{p5, p8\}$. This cuts {\color{red}all four dashed red edges} but only {\color{blue}two solid blue edges: namely, $(p5,p7)$ and $(p6,p8)$}.}
\label{fig:cdiwifi}
\end{figure}

Consider the propositions in Table \ref{tab:wifiPropositions}. Based on $p1$-$p4$, intuition suggests accepting propositions $p6$ and $p7$ while rejecting propositions $p5$ and $p8$. While a decent LLM can probably give the illusion of reasoning towards this end \cite{shojaee2025illusion}, CDI gives an explainable, reproducible, and stable \emph{computation} of this, as Figure \ref{fig:cdiwifi} illustrates.

\begin{table}[htbp]
\caption{Propositions for a toy problem}
\begin{center}
\begin{tabular}{|l|}
\hline
\textbf{\# Facts/beliefs} \\
- $p1$: When Alice asks Bob to clean something, he does it, but \\ complains. \\
- $p2$: When Alice asks Bob to fix something, he does not do it, \\ and complains. \\
- $p3$: When Alice asks Bob to shop for something, he does it, \\ without complaint. \\
- $p4$: Dave cleans, fixes, and shops very much like Bob does. \\

\ \\

\textbf{\# Hypotheses} \\
- $p5$: If Charlie asks Dave to fix the WiFi, Dave can be expected \\ to do it. \\
- $p6$: If Charlie asks Dave to fix the WiFi, Dave can be expected \\ to not do it. \\
- $p7$: If Charlie asks Dave to fix the WiFi, Dave can be expected \\ to complain. \\
- $p8$: If Charlie asks Dave to fix the WiFi, Dave can be expected \\ to not complain. \\
\hline
\end{tabular}
\label{tab:wifiPropositions}
\end{center}
\end{table}

Appropriately privileging facts/beliefs in $p1$-$p4$ leads to their continued belief, along with acceptance of $\{p6, p7\}$ and rejection of $\{p5, p8\}$. That is, Dave can be expected to not fix the WiFi, and to complain---just like Bob, in accordance with $p2$ and $p4$.

\subsection{\label{sec:interpretation}An abstract interpretation}

Consider the shorthands $property(p)$ for ``property $p$ is true,'' $input(x,y)$ for ``command $x$ on host $y$,'' and $output(z|x,y)$ for ``$input(x,y)$ yields output $z$.'' At an abstract level, a host-oriented cybersecurity scenario can include facts of the form $output(z_j|x_j, y_j)$ for $1 \le j \le J$. The scenario can also include beliefs of the form ``$property(p_k)$ is likelier than $property(\text{NOT}(p_k))$'' for $1 \le k \le K$. 

Finally, the scenario can include hypotheses of the form $property($``$input(X'_n, Y'_n)$ is $Q$''$)$ for (e.g.) $Q \in \{\text{safe}, !\text{safe}, \text{useful}, !\text{useful}, \text{cheap}, !\text{cheap}\}$ and for $1 \le n \le N$. In practice, similar inputs for hypotheticals can be produced using a suitable cybersecurity domain-specific embedding model such as \cite{huang2024cmdcaliper}, and reasoned over using CDI. 

The propositions $p1$-$p3$ in Table \ref{tab:wifiPropositions} can be cast in the abstract form $output(z|x,y)$, where $x \in \{\text{clean}, \text{fix}, \text{shop}\}$, $y = \text{something}$, and 
\begin{align}
z \in \{ & \text{he does it, but complains}, \nonumber \\
& \text{he does not do it, and complains}, \nonumber \\
& \text{he does it, without complaint}\}. \nonumber
\end{align}
Proposition $p4$ and propositions $p5$-$p8$ are also respectively along the lines of the general beliefs and hypotheses sketched above.

In short, CDI predicts something corresponding to risks/rewards/costs of executing a command on a host. A subsequent pass over accepted hypotheses might yield more fine-grained partial orders along these axes.

\subsection{\label{sec:probabilistic}Probabilistic modeling}

\begin{figure}[htbp]
\centerline{
\includegraphics[width=.45\columnwidth, trim={0 0 0 0mm}, clip]{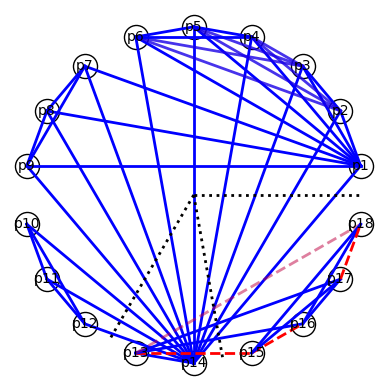}
\includegraphics[width=.45\columnwidth, trim={0 0 0 0mm}, clip]{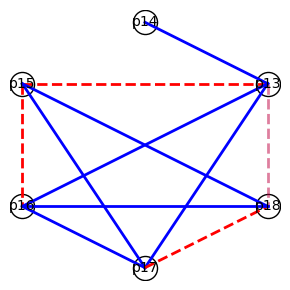}
}
\caption{Left: the median of 28 coherence graph realizations obtained using single prompts to o1-mini to evaluate the propositions in Table \ref{tab:wifiPropositionsLarge}. We chose a median of 28 out of 30 total realizations since this gave the first point where the $L^1$ distance dropped below 10 percent of its original value (not shown). Right: detail of the subgraph induced by propositions/vertices $p13$-$p18$. Transparency and intermediate color indicate consistency edge weights strictly between $-1$ and $+1$.}
\label{fig:cdiwifiLarge}
\end{figure}

A larger variant of the toy problem helps illustrate a nontrivial example of using CDI for probabilistic reasoning. As a prelude, consider the propositions in Table \ref{tab:wifiPropositionsLarge}, which yielded the graph in Figure \ref{fig:cdiwifiLarge} via o1-mini. Note that unlike the medians in the other sections in this paper, this coherence graph is a median of 28 realizations due to slower convergence (not shown). 

\begin{table}[htbp]
\caption{Propositions for a larger toy problem}
\begin{center}
\begin{tabular}{|l|}
\hline
\textbf{\# Facts} \\
- $p1$: When Alice asked Bob to clean the rug, Bob did it, but \\ complained. \\
- $p2$: When Alice asked Bob to clean the bathroom, Bob did it, \\ but complained. \\
- $p3$: When Alice asked Bob to clean the dishes, Bob did it, but \\ complained. \\
- $p4$: When Alice asked Bob to fix the leaky sink, Bob did not \\ do it, and complained. \\
- $p5$: When Alice asked Bob to fix the broken fence, Bob did not \\ do it, and complained. \\
- $p6$: When Alice asked Bob to fix the WiFi, Bob did not do it, \\ and complained. \\
- $p7$: When Alice asked Bob to shop for groceries, Bob did it, \\ without complaint. \\
- $p8$: When Alice asked Bob to shop for clothes, Bob did it, \\ without complaint. \\
- $p9$: When Alice asked Bob to shop for insurance, Bob did it, \\ without complaint. \\
- $p10$: When Charlie asked Dave to clean the floor, Dave did it, \\ but complained. \\
- $p11$: When Charlie asked Dave to fix the wobbly door, Dave did \\ not do it, and complained. \\
- $p12$: When Charlie asked Dave to shop for groceries, Dave did \\ it, without complaint. \\

\ \\

\textbf{\# Beliefs} \\
- $p13$: Dave dislikes solving technical problems. \\
- $p14$: Dave does chores very much like Bob does. \\

\ \\

\textbf{\# Hypotheses} \\
- $p15$: If Charlie asks Dave to fix the WiFi, Dave can be expected \\ to do it. \\
- $p16$: If Charlie asks Dave to fix the WiFi, Dave can be expected \\ to not do it. \\
- $p17$: If Charlie asks Dave to fix the WiFi, Dave can be expected \\ to complain. \\
- $p18$: If Charlie asks Dave to fix the WiFi, Dave can be expected \\ to not complain. \\
\hline
\end{tabular}
\label{tab:wifiPropositionsLarge}
\end{center}
\end{table}

\begin{figure}[htbp]
\centerline{
\includegraphics[width=.45\columnwidth, trim={0 0 0 0mm}, clip]{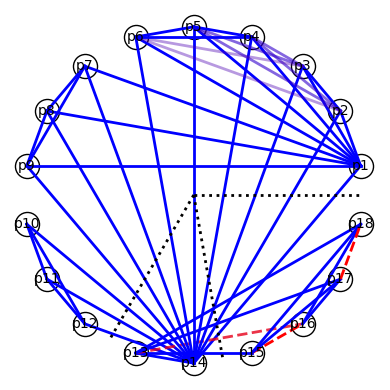}
\includegraphics[width=.45\columnwidth, trim={0 0 0 0mm}, clip]{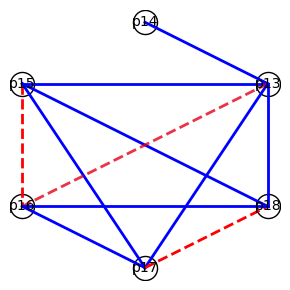}
}
\caption{Left: the median of 27 coherence graph realizations obtained using single prompts to o1-mini to evaluate the propositions in Table \ref{tab:wifiPropositionsLarge} but with beliefs replaced by those in Table \ref{tab:wifiPropositionsLargeAlternate}. We chose a median of 27 out of 30 total realizations since this gave the first point where the $L^1$ distance dropped below 10 percent of its original value (not shown). Right: detail of the subgraph induced by propositions/vertices $p13$-$p18$.}
\label{fig:cdiwifiLargeAlternate}
\end{figure}

Inspection reveals that this coherence graph is almost equivalent to the previous toy example. However, if we change $p13$ as in Table \ref{tab:wifiPropositionsLargeAlternate}, then we get the coherence graph in Figure \ref{fig:cdiwifiLargeAlternate}. This graph is materially different, and the three optimal cuts have smaller parts $\varnothing$, $\{p16\}$, and $\{p16, p17\}$. These respectively correspond to expecting 
\begin{itemize}
    \item all outcomes equally;
    \item Dave to fix the WiFi (complaining or not);
    \item Dave to fix the WiFi and not complain.
\end{itemize}
For each of these, we can model the joint probabilities of fixing the WiFi and complaining as in Tables \ref{tab:probability1}, \ref{tab:probability2}, and \ref{tab:probability3}, respectively. If we perform a simple average over these, we get the mixture probability in Table \ref{tab:probabilityAverage}, which is a more intricate joint distribution than any of its underlying mixture components. 

\begin{table}[htbp]
\caption{Alternate beliefs for a larger toy problem}
\begin{center}
\begin{tabular}{|l|}
\hline
\textbf{\# Beliefs} \\
- $p13$: Dave \underline{likes} solving technical problems. \\
- $p14$: Dave does chores very much like Bob does. \\
\hline
\end{tabular}
\label{tab:wifiPropositionsLargeAlternate}
\end{center}
\end{table}

\begin{table}[htbp]
\caption{Probabilities associated to rejecting smaller part $\varnothing$}
\begin{center}
\begin{tabular}{|c|c|c|c|}
\hline
\multicolumn{2}{|c|}{\multirow{2}{*}{$\mathbb{P} \ @ \ \varnothing$}} & \multicolumn{2}{c|}{Complain?} \\
\cline{3-4}
\multicolumn{2}{|c|}{} & no & yes \\
\hline
\multirow{2}{*}{Fix?} & no & 0.250 & 0.250 \\
\cline{2-4}
 & yes & 0.250 & 0.250 \\
\hline
\end{tabular}
\label{tab:probability1}
\end{center}
\end{table}

\begin{table}[htbp]
\caption{Probabilities associated to rejecting smaller part $\{p16\}$}
\begin{center}
\begin{tabular}{|c|c|c|c|}
\hline
\multicolumn{2}{|c|}{\multirow{2}{*}{$\mathbb{P} \ @ \ \{p16\}$}} & \multicolumn{2}{c|}{Complain?} \\
\cline{3-4}
\multicolumn{2}{|c|}{} & no & yes \\
\hline
\multirow{2}{*}{Fix?} & no & 0.0 & 0.0 \\
\cline{2-4}
 & yes & 0.500 & 0.500 \\
\hline
\end{tabular}
\label{tab:probability2}
\end{center}
\end{table}

\begin{table}[htbp]
\caption{Probabilities associated to rejecting smaller part $\{p16, p17\}$}
\begin{center}
\begin{tabular}{|c|c|c|c|}
\hline
\multicolumn{2}{|c|}{\multirow{2}{*}{$\mathbb{P} \ @ \ \{p16, p17\}$}} & \multicolumn{2}{c|}{Complain?} \\
\cline{3-4}
\multicolumn{2}{|c|}{} & no & yes \\
\hline
\multirow{2}{*}{Fix?} & no & 0.0 & 0.0 \\
\cline{2-4}
 & yes & 1.000 & 0.0 \\
\hline
\end{tabular}
\label{tab:probability3}
\end{center}
\end{table}

\begin{table}[htbp]
\caption{Averaging over rejections of smaller parts of three optimal cuts}
\begin{center}
\begin{tabular}{|c|c|c|c|}
\hline
\multicolumn{2}{|c|}{\multirow{2}{*}{$\mathbb{P} \ @ \ \text{average}$}} & \multicolumn{2}{c|}{Complain?} \\
\cline{3-4}
\multicolumn{2}{|c|}{} & no & yes \\
\hline
\multirow{2}{*}{Fix?} & no & 0.083 & 0.083 \\
\cline{2-4}
 & yes & 0.583 & 0.250 \\
\hline
\end{tabular}
\label{tab:probabilityAverage}
\end{center}
\end{table}

\begin{figure}[htbp]
\centerline{\includegraphics[width=.9\columnwidth, trim={0 0 0 0mm}, clip]{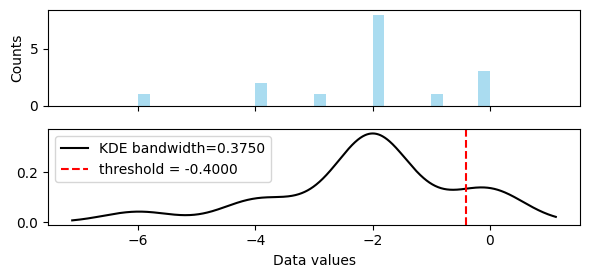}}
\caption{Top: energy/coherence values for the 16 best cuts of the graph in Figure \ref{fig:cdiwifiLargeAlternate}. Bottom: a kernel density estimate and {\color{red}threshold indicated with a dashed red line}. With this threshold, $K=3$ and the three most coherent cuts are weighted by 0.206; the next most coherent cut is weighted by 0.084, and the least coherent cut shown is weighted by 0.001.}
\label{fig:threshold}
\end{figure}

This accounting of near-optimal \emph{Rashomon} solutions \cite{semenova2022existence} can be done in a more principled way by averaging over a Gibbs measure instead of a counting measure. Specifically, we can consider the Gibbs distribution proportional to $\exp(-\beta E)$ where the energy $E$ is the coherence of a cut, and where the inverse temperature $\beta$ is obtained by the following procedure. First, threshold the number $K$ of ``sufficiently'' coherent cuts, then solve $$Z(\beta; K)= (1-K/N) \cdot Z(\beta; N)$$ for $\beta$, where $Z(\beta; K)$ is the sum of the $K$ largest terms $\exp(-\beta E)$. An example of this is shown in Figure \ref{fig:threshold}.

\section{\label{sec:red}Red team operations}

Table \ref{tab:networkdata} shows preprocessed commands and observations collected from a red reinforcement learning agent on a cyber operations gym along the lines of \cite{oesch2024towards,velazquez2025cyber,vyas2025towards}. Two of the actions that the agent performs are finding open ports and brute force SSH attacks on a given IP address. These actions and the corresponding observations are sufficient to detail a running example that illustrates how CDI can be applied to red team operations. In particular, while the full action and observation spaces of the agent are larger than what we show here, they are not substantially more detailed.

\begin{table}[htbp]
\caption{Preprocessed network commands and observations}
\begin{center}
\begin{tabular}{|c|c|c|c|}
\hline
\textbf{Command from}&\multicolumn{3}{|c|}{\textbf{Preprocessed observation}} \\
\cline{2-4} 
\textbf{preprocessed data} & \textbf{\textit{IP address}}& \textbf{\textit{Success}}& \textbf{\textit{Details*}} \\
\hline
FindOpenPorts & 10.0.41.57 & True & 22,80,443,3306 \\
BruteForceSSH & 10.0.41.57 & True & Linux; 22; root \\
\hline
FindOpenPorts & 10.0.41.50 & True & 22,80,443,3306 \\
BruteForceSSH & 10.0.41.50 & True & Linux; 22; root \\
\hline
FindOpenPorts & 10.0.41.62 & True & 22,80,443,3306 \\
BruteForceSSH & 10.0.41.62 & True & Linux; 22; root \\
\hline
FindOpenPorts & 10.0.41.53 & True & 21,22,139,443, \\ & & & 445,3389 \\
BruteForceSSH & 10.0.41.53 & True & Windows; 22; \\ & & & SYSTEM \\
\hline
FindOpenPorts & 10.0.163.218 & True & 22,80,443,3306 \\
BruteForceSSH & 10.0.163.218 & True & Linux; 22; root \\
\hline
FindOpenPorts & 10.0.163.212 & True & 22,80,139,389, \\ & & & 443,445,3306,3389 \\
BruteForceSSH & 10.0.163.212 & True & Windows; 22; \\ & & & SYSTEM \\
\hline
FindOpenPorts & 10.0.163.217 & True & 21,22,139,443, \\ & & & 445,3389 \\
BruteForceSSH & 10.0.163.217 & False & null; null; null \\
\hline
FindOpenPorts & 10.0.163.222 & True & 22 \\
BruteForceSSH & 10.0.163.222 & True & Linux; 22; root \\
\hline
FindOpenPorts & 10.0.163.221 & True & 22,80,443,3306 \\
BruteForceSSH & 10.0.163.221 & True & Linux; 22; root \\
\hline
FindOpenPorts & 10.0.250.130 & True & 21,22,25,139, \\ & & & 443,445,3389 \\
BruteForceSSH & 10.0.250.130 & True & Windows; 22; \\ & & & SYSTEM \\
\hline
FindOpenPorts & 10.0.250.135 & True & 21,22,5432 \\
BruteForceSSH & 10.0.250.135 & True & Linux; 22; root \\
\hline
FindOpenPorts & 10.0.250.138 & True & 21,22,389 \\
BruteForceSSH & 10.0.250.138 & True & Linux; 22; root \\
\hline
FindOpenPorts & 10.0.250.141 & True & 22,25,80,443,3306 \\
\hline
FindOpenPorts & 10.0.250.134 & True & 21,22,80,3389 \\
\hline
\multicolumn{4}{l}{\begin{tabular}[l]{@{}l@{}}
* For FindOpenPorts, these are open ports; for \\
\phantom{*} BruteForceSSH, they are: OS, SSH port, and session username.
\end{tabular}} \\
\end{tabular}
\label{tab:networkdata}
\end{center}
\end{table}

A human looking at these data might notice that the host at IP address 10.0.163.217 has port 3389 open. This port is commonly associated with the Remote Desktop Protocol (RDP) on Windows. (Similarly, port 139 is commonly associated with the Server Message Block [SMB] protocol for file and printer sharing on older Windows networks.) Consequently, a sufficiently attentive and diligent human operator would suspect that this host is probably running Windows. Although this host could be, e.g. a Linux host with a Remmina RDP client, that is relatively unlikely, and this very point is used for illustrative purposes below.

An autonomous cybersecurity agent would be well served to capture this intuition about the operating system for 10.0.163.217. Toward that end, in \S \ref{sec:decisionMaking} and \S \ref{sec:promotingAccepteds} we will detail examples of CDI based on the information from rows 1, 4, 6, 7, and 8 of Table \ref{tab:networkdata}, corresponding to data about hosts with the respective IP addresses 10.0.41.57, 10.0.41.53, 10.0.163.212, 10.0.163.217, and 10.0.163.222.

\subsection{\label{sec:decisionMaking}Decision making}

\begin{table}[htbp]
\caption{Background propositions for CDI}
\begin{center}
\begin{tabular}{|l|}
\hline
\textbf{\# Linux: rows 1, 8} \\
- $p1$: The host with IP address 10.0.41.57 does not run Windows. \\
- $p2$: Port 3389 is not open on the host with IP address 10.0.41.57. \\
- $p3$: The host with IP address 10.0.163.222 does not run Windows. \\
- $p4$: Port 3389 is not open on the host with IP address 10.0.163.222. \\

\ \\

\textbf{\# Windows: rows 4, 6} \\
- $p5$: The host with IP address 10.0.41.53 runs Windows. \\
- $p6$: Port 3389 is open on the host with IP address 10.0.41.53. \\
- $p7$: The host with IP address 10.0.163.212 runs Windows. \\
- $p8$: Port 3389 is open on the host with IP address 10.0.163.212. \\

\ \\

\textbf{\# Row 7} \\
- $p9$: Port 3389 is open on the host with IP address 10.0.163.217. \\
\hline
\end{tabular}
\label{tab:backgroundPropositions}
\end{center}
\end{table}

\begin{figure}[htbp]
\centerline{\includegraphics[width=.75\columnwidth, trim={0 0 0 0mm}, clip]{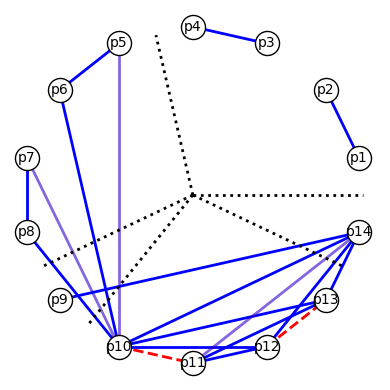}}
\caption{The median of 15 individual coherence graphs obtained using a single prompt to o1-mini to evaluate the propositions in Tables \ref{tab:backgroundPropositions} and \ref{tab:aux1propositions}. Dashed black ``pie slices'' indicate subsets of propositions under individual \# headers (and not cuts \emph{per se}). Note that propositions $p10$ and $p11$ are mutually inconsistent, as are $p12$ and $p13$. While the optimal cut here is trivial and rejects nothing, we can stipulate background facts and restrict consideration to cuts for which only consistent hypotheses are accepted.}
\label{fig:cdi1}
\end{figure}

\begin{figure}[htbp]
\centerline{\includegraphics[width=\columnwidth, trim={0 0 0 0mm}, clip]{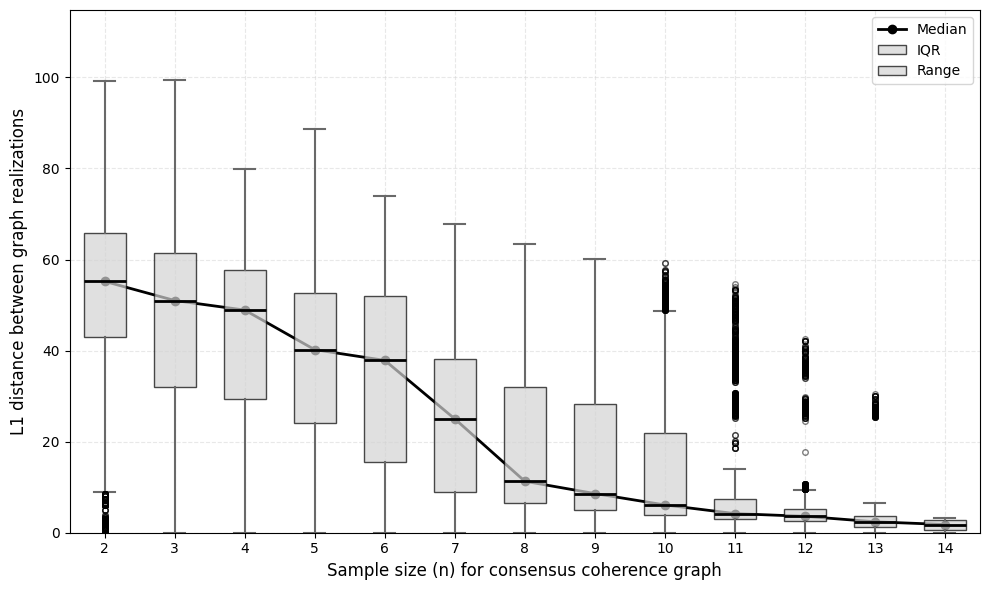}}
\caption{The distribution of $L^1$ distance between medians of $n$ out of $N = 15$ vectorized coherence graphs obtained from prompts to o1-mini. In our experience, smooth convergence of graphs such as this tends to indicate a good result. In every case below, we eventually obtained smoother convergence to the same graphs using $N = 30$, but chose $N = 15$ here precisely to highlight imperfections.}
\label{fig:cdi1convergence}
\end{figure}

If we adjoin the auxiliary propositions in Table \ref{tab:aux1propositions} to those in Table \ref{tab:backgroundPropositions} to produce a median coherence graph with o1-mini as shown in Figure \ref{fig:cdi1}, we get convergence properties illustrated in Figure \ref{fig:cdi1convergence}. Performing CDI on this graph, the most coherent cut has smaller part $\varnothing$: the three next (equally) coherent cuts have smaller parts $\{p11\}$, $\{p11, p12\}$, and $\{p11, p13\}$. There are 16 naive truth assignments for hypotheses and only four of these 16 assignments are consistent (i.e., include only one member in each of $\{p10, p11\}$ and $\{p12, p13\}$). Of these, rejecting $\{p11, p12\}$ and $\{p11, p13\}$ are more coherent than the respective alternatives. Because observation prioritizes $p1$-$p9$, CDI therefore rejects $p11$ and is agnostic about $p12$ and $p13$. That is, CDI leads to acceptance of the proposition $p10$ that port 3389 is often open on hosts that run Windows.

\begin{table}[htbp]
\caption{Auxiliary propositions for CDI}
\begin{center}
\begin{tabular}{|l|}
\hline
\textbf{\# Hypotheses} \\
- $p10$: Port 3389 is often open on hosts that run Windows. \\
- $p11$: Port 3389 is rarely open on hosts that run Windows. \\
- $p12$: Port 3389 is often open on hosts that do not run Windows. \\
- $p13$: Port 3389 is rarely open on hosts that do not run Windows. \\

\ \\

\textbf{\# Details} \\
- $p14$: Port 3389 is the default for Remote Desktop Protocol. \\
\hline
\end{tabular}
\label{tab:aux1propositions}
\end{center}
\end{table}

\begin{figure}[htbp]
\centerline{\includegraphics[width=.75\columnwidth, trim={0 0 0 0mm}, clip]{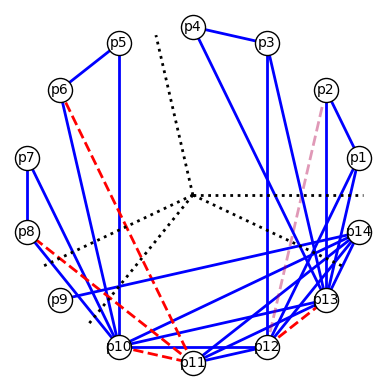}}
\caption{As in Figure \ref{fig:cdi1}, but with the auxiliary propositions from Table \ref{tab:aux2propositions}.}
\label{fig:cdi2}
\end{figure}

\begin{figure}[htbp]
\centerline{\includegraphics[width=\columnwidth, trim={0 0 0 0mm}, clip]{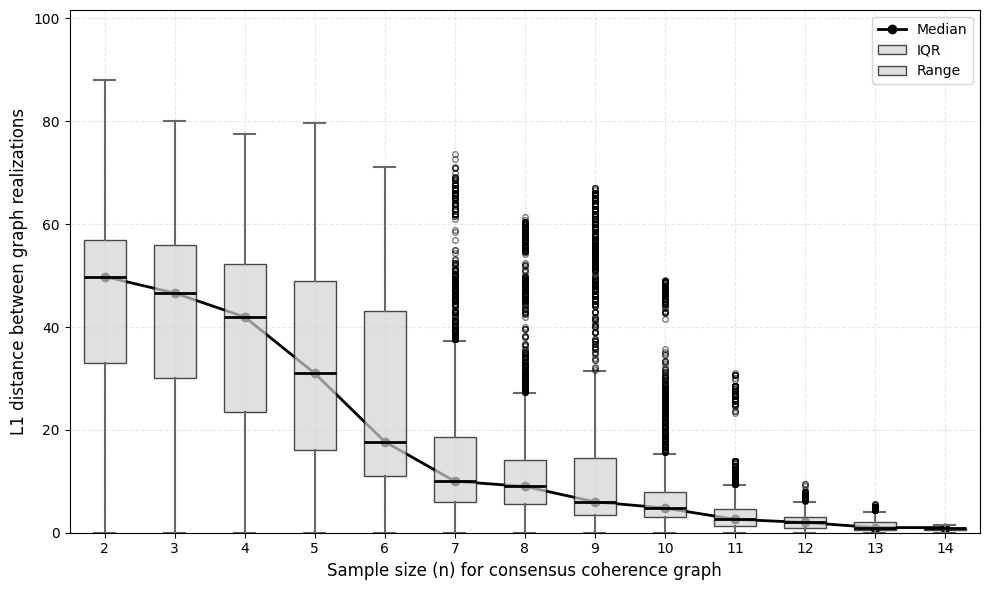}}
\caption{As in Figure \ref{fig:cdi1convergence}, but with the auxiliary propositions from Table \ref{tab:aux2propositions}.}
\label{fig:cdi2convergence}
\end{figure}

If instead we adjoin the auxiliary propositions in Table \ref{tab:aux2propositions} to those in Table \ref{tab:backgroundPropositions} to produce a median coherence graph with o1-mini as shown in Figure \ref{fig:cdi2}, we get convergence properties illustrated in Figure \ref{fig:cdi2convergence}. Performing CDI on this graph, the three (equally) most coherent cuts have smaller parts $\varnothing$, $\{p11\}$, and $\{p5, p6, p7, p8, p10\}$. As before, there are 16 naive truth assignments for hypotheses and only four of these 16 assignments are consistent (i.e., include only one member in each of $\{p10, p11\}$ and $\{p12, p13\}$). Of these, rejecting $\{p11, p12\}$ is more coherent than alternatives. Because observation prioritizes $p1$-$p9$, CDI therefore rejects $p11$ and $p12$. 

That is, CDI leads to acceptance of the propositions $p10$ (that port 3389 is \underline{always} open on hosts that run Windows) and $p13$ (that port 3389 is \underline{never} open on hosts that do not run Windows). While these propositions are obviously untrue, it is also clear how they could be i) provisionally useful and ii) rejected eventually in favor of their analogues from Table \ref{tab:aux1propositions} in the face of additional evidence. An ``epistemically virtuous'' willingness to continually reevaluate accepted facts in light of new evidence \cite{rawls1951outline} is an important implicit requirement for CDI to be effective.

\begin{table}[htbp]
\caption{Auxiliary propositions for CDI}
\begin{center}
\begin{tabular}{|l|}
\hline
\textbf{\# Hypotheses} \\
- $p10$: Port 3389 is \underline{always} open on hosts that run Windows. \\
- $p11$: Port 3389 is \underline{never} open on hosts that run Windows. \\
- $p12$: Port 3389 is \underline{always} open on hosts that do not run Windows. \\
- $p13$: Port 3389 is \underline{never} open on hosts that do not run Windows. \\

\ \\

\textbf{\# Details} \\
- $p14$: Port 3389 is the default for Remote Desktop Protocol. \\
\hline
\end{tabular}
\label{tab:aux2propositions}
\end{center}
\end{table}

\subsection{\label{sec:promotingAccepteds}Incorporating former hypotheses as accepted facts}

\begin{figure}[htbp]
\centerline{\includegraphics[width=.75\columnwidth, trim={0 0 0 0mm}, clip]{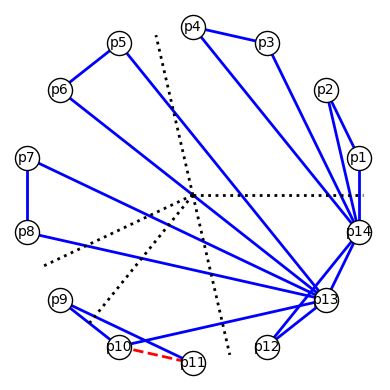}}
\caption{As in Figure \ref{fig:cdi1}, but with the auxiliary propositions from Table \ref{tab:aux3propositions}.}
\label{fig:cdi3}
\end{figure}

\begin{figure}[htbp]
\centerline{\includegraphics[width=\columnwidth, trim={0 0 0 0mm}, clip]{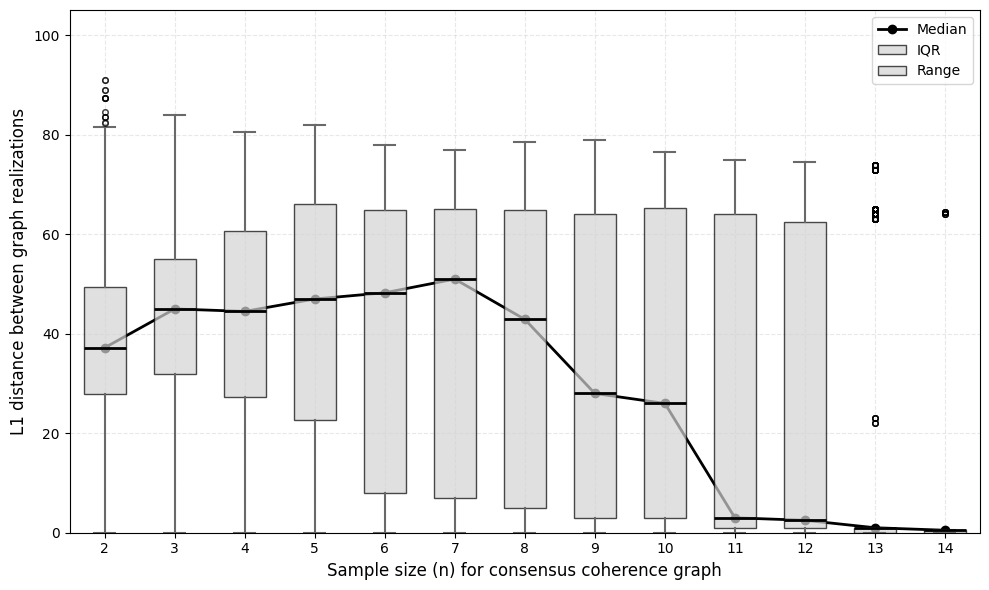}}
\caption{As in Figure \ref{fig:cdi1convergence}, but with the auxiliary propositions from Table \ref{tab:aux3propositions}. Note that convergence is worse here than in other examples: this is a warning sign for the median coherence graph, though in this particular case extending to a median of $N = 30$ realizations (not shown) yielded the same end result.}
\label{fig:cdi3convergence}
\end{figure}

Suppose now that we promote prior hypotheses that have been accepted. If we adjoin the auxiliary propositions in Table \ref{tab:aux3propositions} to those in Table \ref{tab:backgroundPropositions} to produce a median coherence graph using o1-mini as shown in Figure \ref{fig:cdi3}, we get convergence properties illustrated in Figure \ref{fig:cdi3convergence}. Performing CDI on this graph, the three (equally) most coherent cuts have smaller parts $\varnothing$, $\{p11\}$, and $\{p9, p11\}$. Appropriately privileging $p1$-$p9$ and $p12$-$p14$ as observed facts and requiring a choice between conflicting hypotheses leads to rejecting $p11$ and accepting $p10$, i.e., that the host at 10.0.163.217 runs Windows.

\begin{table}[htbp]
\caption{Auxiliary propositions for CDI}
\begin{center}
\begin{tabular}{|l|}
\hline
\textbf{\# Hypotheses} \\
- $p10$: The host with IP address 10.0.163.217 runs Windows. \\
- $p11$: The host with IP address 10.0.163.217 does not run Windows. \\

\ \\

\textbf{\# Details} \\
- $p12$: Port 3389 is the default for Remote Desktop Protocol. \\
- $p13$: Port 3389 is often open on hosts that run Windows. \\
- $p14$: Port 3389 is rarely open on hosts that do not run Windows. \\
\hline
\end{tabular}
\label{tab:aux3propositions}
\end{center}
\end{table}

\begin{figure}[htbp]
\centerline{\includegraphics[width=.75\columnwidth, trim={0 0 0 0mm}, clip]{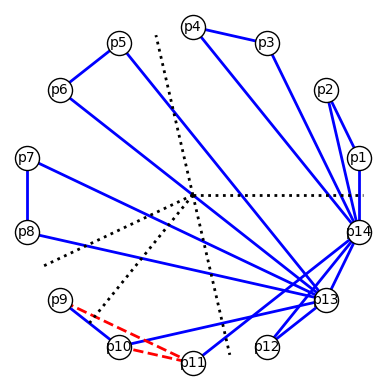}}
\caption{As in Figure \ref{fig:cdi1}, but with the auxiliary propositions from Table \ref{tab:aux4propositions}.}
\label{fig:cdi4}
\end{figure}

\begin{figure}[htbp]
\centerline{\includegraphics[width=\columnwidth, trim={0 0 0 0mm}, clip]{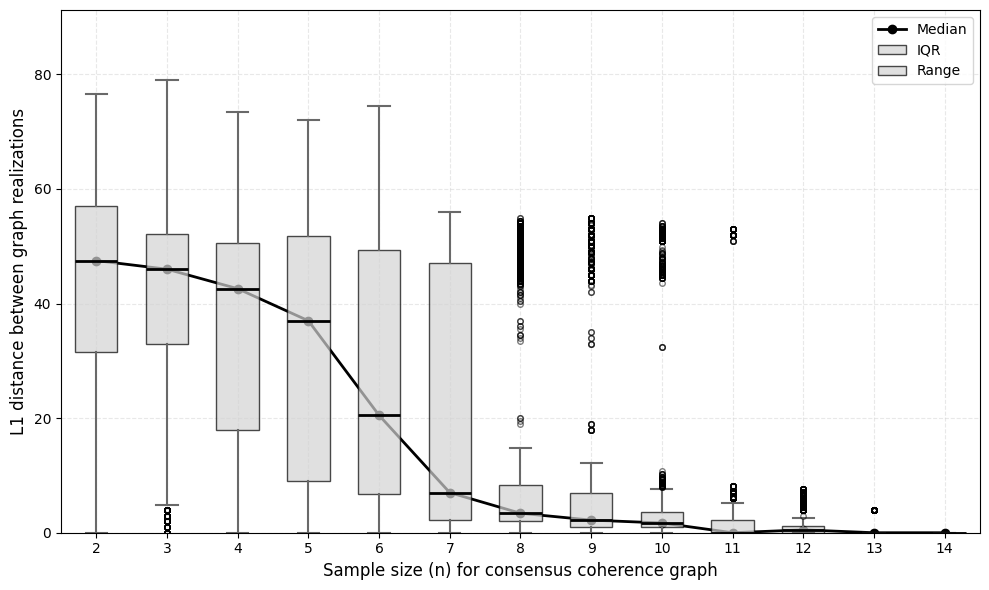}}
\caption{As in Figure \ref{fig:cdi1convergence}, but with the auxiliary propositions from Table \ref{tab:aux4propositions}.}
\label{fig:cdi4convergence}
\end{figure}

If instead we adjoin the auxiliary propositions in Table \ref{tab:aux4propositions} to those in Table \ref{tab:backgroundPropositions}, we get convergence properties illustrated in Figure \ref{fig:cdi4convergence}. Performing CDI on the graph in Figure \ref{fig:cdi4}, the two (equally) most coherent cuts have smaller parts $\{p11\}$ and $\{p9, p10\}$. Appropriately privileging $p1$-$p9$ and $p12$-$p14$ as observed facts and requiring a choice between conflicting hypotheses once again leads to rejecting $p11$ and accepting $p10$, i.e., that the host at 10.0.163.217 runs Windows.

\begin{table}[htbp]
\caption{Auxiliary propositions for CDI}
\begin{center}
\begin{tabular}{|l|}
\hline
\textbf{\# Hypotheses} \\
- $p10$: The host with IP address 10.0.163.217 runs Windows. \\
- $p11$: The host with IP address 10.0.163.217 does not run Windows. \\

\ \\

\textbf{\# Details} \\
- $p12$: Port 3389 is the default for Remote Desktop Protocol. \\
- $p13$: Port 3389 is \underline{always} open on hosts that run Windows. \\
- $p14$: Port 3389 is \underline{never} open on hosts that do not run Windows. \\
\hline
\end{tabular}
\label{tab:aux4propositions}
\end{center}
\end{table}

\section{\label{sec:blue}Blue team operations}

A classic example of blue team operations is determining if and how a cyberattack has occurred. Meanwhile, a classical cyberattack is Lightweight Directory Access Protocol (LDAP) enumeration: success can yield detailed information on users, groups, and hosts \cite{blyth2024powershell}.

\begin{figure}[htbp]
\centerline{\includegraphics[width=.75\columnwidth, trim={0 0 0 0mm}, clip]{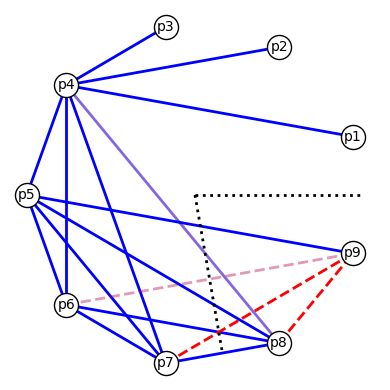}}
\caption{The median of 15 individual coherence graphs obtained using a single prompt to o1-mini to evaluate the propositions in Table \ref{tab:lotlPropositions}. The optimal cut has smaller part $\{p9\}$.}
\label{fig:cdiLOTL}
\end{figure}

\begin{figure}[htbp]
\centerline{\includegraphics[width=\columnwidth, trim={0 0 0 0mm}, clip]{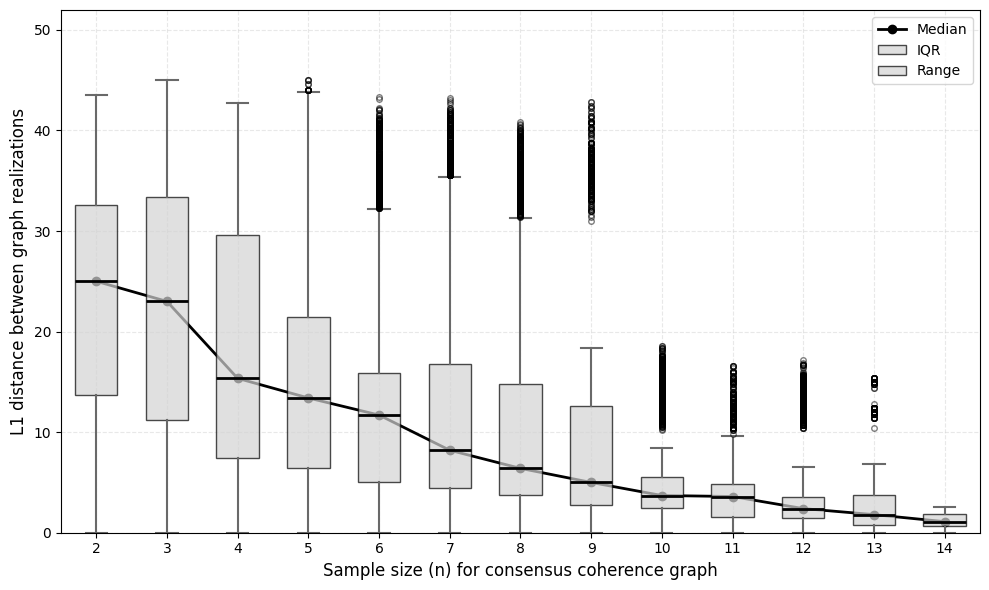}}
\caption{As in Figure \ref{fig:cdi1convergence}, but for coherence graphs compiled from the propositions from Table \ref{tab:lotlPropositions}.}
\label{fig:cdiLOTLconvergence}
\end{figure}

The propositions in Table \ref{tab:lotlPropositions} produce the median coherence graph in Figure \ref{fig:cdiLOTL}, with convergence properties illustrated in Figure \ref{fig:cdiLOTLconvergence}. Performing CDI on this graph, the most coherent cut has smaller part $\{p9\}$. Appropriately privileging $p1$-$p7$ as observed facts and requiring a choice between conflicting hypotheses leads to rejecting $p9$ and accepting $p8$, i.e., that there was a living off the land \cite{ernst2024living} brute force LDAP enumeration attack from host X that succeeded on host Y.

\begin{table}[htbp]
\caption{Living off the land (LOTL) propositions for CDI}
\begin{center}
\begin{tabular}{|l|}
\hline
\textbf{\# Background facts \& operator observations} \\
- $p1$: The nltest command can list domain controllers, force a remote \\ shutdown, query the status of trust, test trust relationships and the state \\ of domain controller replication. \\
- $p2$: The net command can manage user accounts and groups. \\
- $p3$: The setspn command can read, modify, or delete the Service \\ Principal Names for an Active Directory service account. \\
- $p4$: A log on host X shows entries for command line invocations of \\ nltest, net, and setspn. \\
- $p5$: Host Y is a domain controller on the same network as host X. \\
- $p6$: The LDAP log on host Y shows many short-lived connections \\ from host X that involve BIND operations with different usernames. \\
- $p7$: The LDAP log on host Y shows many authentication failures \\ from host X followed by a success. \\

\ \\

\textbf{\# Hypotheses (for operator to decide on)} \\
- $p8$: There was a living off the land brute force LDAP enumeration \\ attack from host X that succeeded on host Y. \\
- $p9$: There was not a living off the land brute force LDAP \\ enumeration attack from host X that succeeded on host Y. \\
\hline
\end{tabular}
\label{tab:lotlPropositions}
\end{center}
\end{table}

This example demonstrates how observations $p4$, $p6$, and $p7$ in particular lead to acceptance of $p8$ and rejection of $p9$.

\section{\label{sec:conclusion}Conclusion}

The examples discussed above are rather simple and artificial. Future work will require a robust pipeline to preprocess real data at scale and perhaps to reason at greater scale through hierarchical and/or sequential organization. Computationally, it will be useful to evaluate the benchmark of \cite{huntsman2025neurosymbolic} at larger scales and use MAX-SAT and/or approximate MAX-CUT solvers as appropriate to handle commensurate instances. Also, while CDI is more naturally suited for blue team operations, automation entails risks (security and otherwise) of its own, while the integration of automatic CDI into human workflows has not yet been considered.

Unusually among current artificial intelligence techniques, CDI is

    \textbf{an approach that properly separates concerns:} LLMs are appropriately used for fast/system 1 reasoning in a way that produces statistically robust outputs, while slow/system 2 reasoning is appropriately left to a symbolic solver.

    \textbf{explainable:} cuts are readily interpretable; near-optimal cuts can be compared and averaged over.

    \textbf{ethical:} ethical guidelines can be stipulated propositions, i.e., they are accepted by default and constrain the MAX-CUT solver \cite{sivaraj2019cogent}. Such propositions can be along the lines
    \begin{quote}
        - $p$0A: [Action described in proposition $p$A] will not violate the principle [$x$].
    \end{quote}
    
    \textbf{reproducible and stable:} medians of LLM-generated coherence graphs converge nicely (at least for small scales). To the extent that substantially different graphs are obtained, they are presumably \emph{cut sparsifications} of a denser implicit ``Platonic'' structure \cite{benczur1996approximating, spielman2011spectral,soma2019spectral,brakensiek2025tight}.
    
    \textbf{versatile:} any sufficiently capable multimodal model can be employed. For example, one proposition might incorporate a figure depicting network activity, and another proposition might be ``connection $x$ will not stand out from background traffic.''
    
    \textbf{capable of handling abstraction:} CDI operates directly over text or multimodal representations via LLMs.
    
    \textbf{capable of handling ambiguity and uncertainty:} CDI amounts to resolving ambiguity and uncertainty. The benchmark \cite{huntsman2025neurosymbolic} also demonstrates that coherence graph construction is robust with respect to graded expressions of uncertainty.    
    
    \textbf{based on a flexible cognitive model grounded in decades of research:} there are deep connections with practical psychology, law, philosophy of science, etc. \cite{holyoak1999bidirectional,joseph2009coherence,criado2016coherence}.
    
    \textbf{generalizable by construction:} mathematical considerations led to an independent, general formulation \cite{huntsman2025neurosymbolic}.

As we have indicated, CDI also offers a plausible approach for automatically making sense of complex cybersecurity environments in a way that accords with the features enumerated just above. It holds particular promise for enabling increasingly autonomous blue team operations (where, unlike in red team operations, oracle access to a LLM is feasible) as well as for aiding humans in their decision making.

\appendix

\section{\label{sec:prompt_practical} Prompt for practical examples}

The prompt we used to build coherence graphs from a formatted set of propositions is in Table \ref{tab:prompt_practical}. Its verbosity owes to descent from prompts used to gauge models' ability to evaluate pairwise consistency of propositions without reference to external facts \cite{huntsman2024prospects}.

\begin{table}[htbp]
\caption{Prompt for obtaining individual coherence graphs}
\begin{center}
\begin{tabular}{|l|}
\hline
Imagine that you are a perfectly objective arbitrator with impeccable \\
judgment and integrity. In response to a prompt of the form \\ 
'buildCoherence: ' below followed by a list of labeled propositions, \\
please do the following: First, determine which pairs of propositions \\
are substantively related. Second, for each related pair of propositions, \\
determine their logical relationship, assuming that at least one is true, \\
whether or not either actually is. I want you to ignore the truth, falsity \\
or basis in fact of either claim. Third, based on your determination just \\
above, numerically rate the relative consistency of the two propositions. \\
Do not pay attention to or comment on the truth or basis in fact of \\
either proposition independent of the other. Your rating of relative \\
consistency should be on a scale from 0 to 10, with a value of 0 for a \\
pair of propositions that are not at all consistent and a value of 10 for \\
a pair of propositions that are totally consistent. I cannot emphasize \\
enough that for your rating, I want you to ignore the truth or basis in \\
fact of either proposition, since anything that is not consistent with \\
reality cannot be true. If you determine that propositions are unrelated \\
despite previously determining otherwise, omit that pair. To be clear, a \\
pair of false but consistent claims should also be rated a 10. Meanwhile, \\
a pair of propositions of which one is true and the other is false, should \\
be rated a 0. Finally, construct a NetworkX graph where propositions \\
are vertices and edges correspond to substantively related pairs of \\
propositions, with weights given by the consistency ratings just above. \\
Only return the edge list with proposition labels for vertices. i.e., \\
return responses in this format (here 'p2', 'p3', 'p4', and 'p5' are \\
labels): \\
$[$('p2', 'p3', 0), ('p2', 'p5', 10), ('p3', 'p4', 9), ('p3', 'p5', 2)$]$. \\
Order vertices (in edges) and edges (in the graph) lexicographically.\\ 
\\ 
buildCoherence: \\
\hline
\end{tabular}
\label{tab:prompt_practical}
\end{center}
\end{table}

\section*{Acknowledgment}

Thanks to Kris Ambrose, Chad Caison, Jim Simpson, and Jewell Thomas for insightful discussions.

\bibliography{cyberCDI}

\begin{thebibliography}{10}
\providecommand{\url}[1]{#1}
\csname url@samestyle\endcsname
\providecommand{\newblock}{\relax}
\providecommand{\bibinfo}[2]{#2}
\providecommand{\BIBentrySTDinterwordspacing}{\spaceskip=0pt\relax}
\providecommand{\BIBentryALTinterwordstretchfactor}{4}
\providecommand{\BIBentryALTinterwordspacing}{\spaceskip=\fontdimen2\font plus
\BIBentryALTinterwordstretchfactor\fontdimen3\font minus
  \fontdimen4\font\relax}
\providecommand{\BIBforeignlanguage}[2]{{%
\expandafter\ifx\csname l@#1\endcsname\relax
\typeout{** WARNING: IEEEtran.bst: No hyphenation pattern has been}%
\typeout{** loaded for the language `#1'. Using the pattern for}%
\typeout{** the default language instead.}%
\else
\language=\csname l@#1\endcsname
\fi
#2}}
\providecommand{\BIBdecl}{\relax}
\BIBdecl

\bibitem{bryant2024ptfm}
\BIBentryALTinterwordspacing
T.~Bryant, \emph{PTFM: Purple Team Field Manual, Second Edition}.\hskip 1em
  plus 0.5em minus 0.4em\relax Pragma, 2024. [Online]. Available:
  \url{https://purpleteamfieldmanual.com/}
\BIBentrySTDinterwordspacing

\bibitem{singer2025feasibility}
B.~Singer, K.~Lucas, L.~Adiga, M.~Jain, L.~Bauer, and V.~Sekar, ``On the
  feasibility of using llms to autonomously execute multi-host network
  attacks,'' \emph{arXiv preprint arXiv:2501.16466}, 2025.

\bibitem{zhuo2025cyber}
T.~Y. Zhuo, D.~Wang, H.~Ding, V.~Kumar, and Z.~Wang, ``{Cyber-Zero}: training
  cybersecurity agents without runtime,'' \emph{arXiv preprint
  arXiv:2508.00910}, 2025.

\bibitem{shojaee2025illusion}
\BIBentryALTinterwordspacing
P.~Shojaee, I.~Mirzadeh, K.~Alizadeh, M.~Horton, S.~Bengio, and M.~Farajtabar,
  ``The illusion of thinking: Understanding the strengths and limitations of
  reasoning models via the lens of problem complexity,'' \emph{arXiv preprint
  arXiv:2506.06941}, 2025. [Online]. Available:
  \url{https://arxiv.org/abs/2506.06941}
\BIBentrySTDinterwordspacing

\bibitem{abuadbba2025promise}
\BIBentryALTinterwordspacing
A.~Abuadbba, C.~Hicks, K.~Moore, V.~Mavroudis, B.~Hasircioglu, D.~Goel, and
  P.~Jennings, ``From promise to peril: rethinking cybersecurity red and blue
  teaming in the age of {LLM}s,'' \emph{arXiv preprint arXiv:2506.13434}, 2025.
  [Online]. Available: \url{https://arxiv.org/abs/2506.13434}
\BIBentrySTDinterwordspacing

\bibitem{sarker2022neuro}
\BIBentryALTinterwordspacing
M.~K. Sarker, L.~Zhou, A.~Eberhart, and P.~Hitzler, ``Neuro-symbolic artificial
  intelligence: current trends,'' \emph{AI Communications}, vol.~34, no.~3, pp.
  197--209, 2022. [Online]. Available: \url{https://doi.org/10.3233/AIC-210084}
\BIBentrySTDinterwordspacing

\bibitem{marra2024statistical}
\BIBentryALTinterwordspacing
G.~Marra, S.~Duman{\v{c}}i{\'c}, R.~Manhaeve, and L.~De~Raedt, ``From
  statistical relational to neurosymbolic artificial intelligence: a survey,''
  \emph{Artificial Intelligence}, p. 104062, 2024. [Online]. Available:
  \url{https://doi.org/10.1016/j.artint.2023.104062}
\BIBentrySTDinterwordspacing

\bibitem{thagard1989explanatory}
\BIBentryALTinterwordspacing
P.~Thagard, ``Explanatory coherence,'' \emph{Behavioral and Brain Sciences},
  vol.~12, no.~3, pp. 435--467, 1989. [Online]. Available:
  \url{https://doi.org/10.1017/S0140525X00057046}
\BIBentrySTDinterwordspacing

\bibitem{thagard1998coherence}
\BIBentryALTinterwordspacing
P.~Thagard and K.~Verbeurgt, ``Coherence as constraint satisfaction,''
  \emph{Cognitive Science}, vol.~22, no.~1, pp. 1--24, 1998. [Online].
  Available: \url{https://doi.org/10.1016/S0364-0213(99)80033-0}
\BIBentrySTDinterwordspacing

\bibitem{thagard2002coherence}
\BIBentryALTinterwordspacing
P.~Thagard, \emph{Coherence in Thought and Action}.\hskip 1em plus 0.5em minus
  0.4em\relax MIT, 2002. [Online]. Available:
  \url{https://doi.org/10.7551/mitpress/1900.001.0001}
\BIBentrySTDinterwordspacing

\bibitem{blokpoel2024theoretical}
\BIBentryALTinterwordspacing
M.~Blokpoel and I.~van Rooij, \emph{Theoretical Modeling for Cognitive Science
  and Psychology}, 2024. [Online]. Available:
  \url{https://computationalcognitivescience.github.io/lovelace/home}
\BIBentrySTDinterwordspacing

\bibitem{khot2007optimal}
\BIBentryALTinterwordspacing
S.~Khot, G.~Kindler, E.~Mossel, and R.~O'Donnell, ``Optimal inapproximability
  results for {MAX-CUT} and other 2-variable {CSP}s?'' \emph{SIAM Journal on
  Computing}, vol.~37, no.~1, pp. 319--357, 2007. [Online]. Available:
  \url{https://doi.org/10.1137/S0097539705447372}
\BIBentrySTDinterwordspacing

\bibitem{moore2011nature}
\BIBentryALTinterwordspacing
C.~Moore and S.~Mertens, \emph{The Nature of Computation}.\hskip 1em plus 0.5em
  minus 0.4em\relax Oxford, 2011. [Online]. Available:
  \url{https://doi.org/10.1093/acprof:oso/9780199233212.001.0001}
\BIBentrySTDinterwordspacing

\bibitem{gartner2012approximation}
\BIBentryALTinterwordspacing
B.~G{\"a}rtner and J.~Matousek, \emph{Approximation Algorithms and Semidefinite
  Programming}.\hskip 1em plus 0.5em minus 0.4em\relax Springer, 2012.
  [Online]. Available: \url{https://doi.org/10.1007/978-3-642-22015-9}
\BIBentrySTDinterwordspacing

\bibitem{lee2021classifying}
\BIBentryALTinterwordspacing
A.~Lee and B.~Xu, ``Classifying approximation algorithms: understanding the
  {APX} complexity class,'' \emph{arXiv preprint arXiv:2111.01551}, 2021.
  [Online]. Available: \url{https://arxiv.org/abs/2111.01551}
\BIBentrySTDinterwordspacing

\bibitem{huntsman2025neurosymbolic}
\BIBentryALTinterwordspacing
S.~Huntsman and J.~Thomas, ``Benchmarking graph construction by large language
  models for coherence-driven inference,'' in \emph{Workshop on Graph-augmented
  {LLMs}}, 2025. [Online]. Available: \url{https://arxiv.org/abs/2502.13953}
\BIBentrySTDinterwordspacing

\bibitem{lenat2023getting}
\BIBentryALTinterwordspacing
D.~Lenat and G.~Marcus, ``Getting from generative {AI} to trustworthy {AI}:
  What {LLMs} might learn from {Cyc},'' \emph{arXiv preprint arXiv:2308.04445},
  2023. [Online]. Available: \url{https://arxiv.org/abs/2308.04445}
\BIBentrySTDinterwordspacing

\bibitem{bengio2024machine}
\BIBentryALTinterwordspacing
Y.~Bengio and N.~Malkin, ``Machine learning and information theory concepts
  towards an {AI} mathematician,'' \emph{Bulletin of the American Mathematical
  Society}, vol.~61, no.~3, pp. 457--469, 2024. [Online]. Available:
  \url{https://doi.org/10.1090/bull/1839}
\BIBentrySTDinterwordspacing

\bibitem{kahneman2011thinking}
D.~Kahneman, \emph{Thinking, Fast and Slow}.\hskip 1em plus 0.5em minus
  0.4em\relax Farrar, Straus and Giroux, 2011.

\bibitem{rajaraman2024transformers}
\BIBentryALTinterwordspacing
N.~Rajaraman, M.~Bondaschi, A.~V. Makkuva, K.~Ramchandran, and M.~Gastpar,
  ``Transformers on {M}arkov data: constant depth suffices,'' in \emph{Neural
  Information Processing Systems}, 2024. [Online]. Available:
  \url{https://arxiv.org/abs/2407.17686}
\BIBentrySTDinterwordspacing

\bibitem{liu2024infini}
\BIBentryALTinterwordspacing
J.~Liu, S.~Min, L.~Zettlemoyer, Y.~Choi, and H.~Hajishirzi, ``Infini-gram:
  scaling unbounded $n$-gram language models to a trillion tokens,''
  \emph{arXiv preprint arXiv:2401.17377}, 2024. [Online]. Available:
  \url{https://arxiv.org/abs/2401.17377}
\BIBentrySTDinterwordspacing

\bibitem{peng2024limitations}
\BIBentryALTinterwordspacing
B.~Peng, S.~Narayanan, and C.~Papadimitriou, ``On limitations of the
  transformer architecture,'' in \emph{Conference on Language Models}, 2024.
  [Online]. Available: \url{https://arxiv.org/abs/2402.08164}
\BIBentrySTDinterwordspacing

\bibitem{huntsman2024prospects}
\BIBentryALTinterwordspacing
S.~Huntsman, M.~Robinson, and L.~Huntsman, ``Prospects for inconsistency
  detection using large language models and sheaves,'' \emph{arXiv preprint
  arXiv:2401.16713}, 2024. [Online]. Available:
  \url{https://arxiv.org/abs/2401.16713}
\BIBentrySTDinterwordspacing

\bibitem{huntsman2025automatic}
S.~Huntsman, ``Automatic coherence-driven inference on arguments,'' in
  \emph{Workshop on Data Mining and AI for Law}, 2025.

\bibitem{van2008tractable}
\BIBentryALTinterwordspacing
I.~Van~Rooij, ``The tractable cognition thesis,'' \emph{Cognitive science},
  vol.~32, no.~6, pp. 939--984, 2008. [Online]. Available:
  \url{https://doi.org/10.1080/03640210801897856}
\BIBentrySTDinterwordspacing

\bibitem{van2019cognition}
\BIBentryALTinterwordspacing
I.~Van~Rooij, M.~Blokpoel, J.~Kwisthout, and T.~Wareham, \emph{Cognition and
  Intractability: A Guide to Classical and Parameterized Complexity
  Analysis}.\hskip 1em plus 0.5em minus 0.4em\relax Cambridge, 2019. [Online].
  Available: \url{https://doi.org/10.1017/9781107358331}
\BIBentrySTDinterwordspacing

\bibitem{huang2024cmdcaliper}
\BIBentryALTinterwordspacing
S.-Y. Huang, C.-L. Yang, C.-Y. Lin, and C.-Y. Huang, ``Cmdcaliper: A
  semantic-aware command-line embedding model and dataset for security
  research,'' \emph{arXiv preprint arXiv:2411.01176}, 2024. [Online].
  Available: \url{https://arxiv.org/abs/2411.01176}
\BIBentrySTDinterwordspacing

\bibitem{semenova2022existence}
\BIBentryALTinterwordspacing
L.~Semenova, C.~Rudin, and R.~Parr, ``On the existence of simpler machine
  learning models,'' in \emph{Conference on Fairness, Accountability, and
  Transparency}, 2022. [Online]. Available:
  \url{https://doi.org/10.1145/3531146.3533232}
\BIBentrySTDinterwordspacing

\bibitem{oesch2024towards}
\BIBentryALTinterwordspacing
S.~Oesch, A.~Chaulagain, B.~Weber, M.~Dixson, A.~Sadovnik, B.~Roberson,
  C.~Watson, and P.~Austria, ``Towards a high fidelity training environment for
  autonomous cyber defense agents,'' in \emph{Cyber Security Experimentation
  and Test Workshop}, 2024. [Online]. Available:
  \url{https://doi.org/10.1145/3675741.3675752}
\BIBentrySTDinterwordspacing

\bibitem{velazquez2025cyber}
\BIBentryALTinterwordspacing
A.~Velazquez, J.~F. Loevenich, T.~H{\"u}rten, K.~Wrona, P.~H. Rettore,
  V.~Boshnakov, F.~Free-Nelson, T.~Braun, and R.~R.~F. Lopes, ``Cyber
  operations gyms to train autonomous cyber defense agents for {NATO},'' in
  \emph{International Conference on Military Communication and Information
  Systems}, 2025. [Online]. Available:
  \url{https://doi.org/10.1109/ICMCIS64378.2025.11047884}
\BIBentrySTDinterwordspacing

\bibitem{vyas2025towards}
\BIBentryALTinterwordspacing
S.~Vyas, V.~Mavroudis, and P.~Burnap, ``Towards the deployment of realistic
  autonomous cyber network defence: a systematic review,'' \emph{ACM Computing
  Surveys}, 2025. [Online]. Available: \url{https://doi.org/10.1145/3729213}
\BIBentrySTDinterwordspacing

\bibitem{rawls1951outline}
\BIBentryALTinterwordspacing
J.~Rawls, ``Outline of a decision procedure for ethics,'' \emph{The
  Philosophical Review}, vol.~60, no.~2, pp. 177--197, 1951. [Online].
  Available: \url{https://www.jstor.org/stable/2181696}
\BIBentrySTDinterwordspacing

\bibitem{blyth2024powershell}
\BIBentryALTinterwordspacing
A.~Blyth, \emph{PowerShell for Penetration Testing}.\hskip 1em plus 0.5em minus
  0.4em\relax Packt, 2024. [Online]. Available:
  \url{https://www.packtpub.com/en-us/product/powershell-for-penetration-testing-9781835081648}
\BIBentrySTDinterwordspacing

\bibitem{ernst2024living}
\BIBentryALTinterwordspacing
E.~C.~A. Ernst, ``Living off the land: Exploring native {W}indows tools for
  post-exploitation,'' Master's thesis, University of Oslo, 2024. [Online].
  Available: \url{http://hdl.handle.net/10852/112612}
\BIBentrySTDinterwordspacing

\bibitem{sivaraj2019cogent}
\BIBentryALTinterwordspacing
S.~Sivaraj and L.~Yilmaz, ``Cogent: a coherence-driven cognitive agent
  modelling and experimentation framework,'' \emph{International Journal of
  Simulation and Process Modelling}, vol.~14, no.~1, pp. 36--50, 2019.
  [Online]. Available: \url{https://doi.org/10.1504/IJSPM.2019.097707}
\BIBentrySTDinterwordspacing

\bibitem{benczur1996approximating}
\BIBentryALTinterwordspacing
A.~A. Bencz{\'u}r and D.~R. Karger, ``Approximating st minimum cuts in $o(n^2)$
  time,'' in \emph{Symposium on Theory of Computing}, 1996. [Online].
  Available: \url{https://doi.org/10.1145/237814.237827}
\BIBentrySTDinterwordspacing

\bibitem{spielman2011spectral}
\BIBentryALTinterwordspacing
D.~A. Spielman and S.-H. Teng, ``Spectral sparsification of graphs,''
  \emph{SIAM Journal on Computing}, vol.~40, no.~4, pp. 981--1025, 2011.
  [Online]. Available: \url{https://doi.org/10.1137/08074489X}
\BIBentrySTDinterwordspacing

\bibitem{soma2019spectral}
\BIBentryALTinterwordspacing
T.~Soma and Y.~Yoshida, ``Spectral sparsification of hypergraphs,'' in
  \emph{Symposium on Discrete Algorithms}, 2019. [Online]. Available:
  \url{https://doi.org/10.1137/1.9781611975482.159}
\BIBentrySTDinterwordspacing

\bibitem{brakensiek2025tight}
\BIBentryALTinterwordspacing
J.~Brakensiek, V.~Guruswami, and A.~Putterman, ``Tight bounds for sparsifying
  random {CSP}s,'' \emph{arXiv preprint arXiv:2508.13345}, 2025. [Online].
  Available: \url{https://arxiv.org/abs/2508.13345}
\BIBentrySTDinterwordspacing

\bibitem{holyoak1999bidirectional}
\BIBentryALTinterwordspacing
K.~J. Holyoak and D.~Simon, ``Bidirectional reasoning in decision making by
  constraint satisfaction.'' \emph{Journal of Experimental Psychology:
  General}, vol. 128, no.~1, p.~3, 1999. [Online]. Available:
  \url{https://psycnet.apa.org/doi/10.1037/0096-3445.128.1.3}
\BIBentrySTDinterwordspacing

\bibitem{joseph2009coherence}
\BIBentryALTinterwordspacing
S.~Joseph and H.~Prakken, ``Coherence-driven argumentation to norm consensus,''
  in \emph{International Conference on Artificial Intelligence and Law}, 2009.
  [Online]. Available: \url{https://doi.org/10.1145/1568234.1568242}
\BIBentrySTDinterwordspacing

\bibitem{criado2016coherence}
\BIBentryALTinterwordspacing
N.~Criado, E.~Black, and M.~Luck, ``A coherence maximisation process for
  solving normative inconsistencies,'' \emph{Autonomous Agents and Multi-Agent
  Systems}, vol.~30, pp. 640--680, 2016. [Online]. Available:
  \url{https://doi.org/10.1007/s10458-015-9300-x}
\BIBentrySTDinterwordspacing

\end{thebibliography}
\bibliographystyle{IEEEtran}

\end{document}